\def\acf{\rm ACF}

\def\modul{\hskip 2pt {\rm mod}\hskip 2pt}

\def\qed{\hfill \mbox{\raggedright \rule{0.09in}{0.09in}}}

\def\qed{\hfill \mbox{\raggedright \rule{0.09in}{0.09in}}}

 \documentclass{article}

  \usepackage{setspace}
  \usepackage{graphicx}
  \usepackage{amsmath}
     \usepackage{fullpage}

 \usepackage{amssymb}
\title{Doppler Tolerance, Complementary Code Sets and
Generalized Thue-Morse Sequences}

\author{ Hieu D. Nguyen\\
Department of Mathematics, Rowan University, Glassboro, NJ\\
Gregory E. Coxson\\
Naval Research Laboratory, Washington DC}

\date{10-10-2014}
  
\begin{document}

    \maketitle

 \begin{abstract}
We generalize the construction of Doppler-tolerant Golay complementary waveforms by 
Pezeshki-Calderbank-Moran-Howard to complementary code sets having more than two codes. 
This is accomplished by exploiting number-theoretic results involving the sum-of-digits
function, equal sums of like powers, and a generalization to more than two symbols of the
classical two-symbol Prouhet-Thue-Morse sequence.

\end{abstract}

\vskip 6pt


 \section{Introduction}
 
A set of $K$ unimodular codes of length $N$ is complementary if corresponding sidelobes of the
autocorrelations of the separate codes sum to zero.    These sets find uses in waveform design for
enhanced detection in radar systems [1] and in communication systems [2][3].     When the code is 
binary, the set is called a Golay complementary pair, after Marcel Golay who discovered these sets 
while solving a problem in infrared spectrometry [4].  Complementary Code Matrices (CCMs) 
provide a useful matrix formulation for the study of complementary code sets [5].   
Given a set of $K$ codes of length $N$, the corresponding $N\times K$ complementary code 
matrix (CCM) has the $k^{th}$ code as its $k^{th}$ column, $k=1,\ldots,K$.

Complementary code sets have yet to be widely used for radar waveform designs due to certain design challenges.  
These include sensitivity to Doppler shift due to non-zero relative velocity of a target relative to the radar platform [1][6].    
Complementary code sets may be used in a number of ways in waveform design.
Two of these ways are the time-separation approach, where time-separated pulses
or subpulses are phase coded using different codes in the set [7][8], and frequency-separation,
where the different codes are used for phase encoding of separate components of a signal 
and are transmitted concurrently using pulses with different center frequencies [9][10].   The time-separation 
approach is especially sensitive to Doppler shift.

With time-separated pulses encoded using the codes from a complementary set, pulse returns may be 
match filtered separately and then added to give zero autocorrelation sidelobes, in theory, a 
desirable result for radar detection.   However, target relative velocity yields 
a phase shift pulse to pulse, and therefore a phase shift of sidelobes, thus preventing zero 
sidelobe sums in general.  

The development in this paper builds on work by Pezeshki, Howard, Moran, Calderbank, Chi 
and Searle [4][7][8][9][10].  In particular, in [7], Chi, Calderbank and Pezeshki  
consider pulse trains in which the pulses are phase coded with binary codes in
a Golay complementary pair.  They show that for any given $M$, $M^{th}$-order nulls can be created
about the zero-Doppler axis of the ambiguity function by mapping the codes to pulses in an order
specified by the well-known (two-symbol) Prouhet-Thue-Morse sequence.  We show that the result may 
be generalized to $(N,K)$ complementary code sets for $K \ge 2$  by using a generalized
Prouhet-Thue-Morse sequence using $m\ge 2$ symbols.  The approach also makes use of
results related to the Tarry-Escott problem [11][15][16], and related number-theoretic entities such as the 
digit-sum function [12] and equal sums of like powers [17].  Finally, it is shown that the transmission period and the total number of pulses transmitted may be reduced by using multiple antennas to transmit separate pulse trains staggered in time.

 \section{Notations and Terminology}

\noindent
{\bf Definition 2.1}.  A $p$-{\em phase} matrix $Q$ is one whose entries are $p$-th roots of unity, i.e. roots of $z^p=1$.

\vskip 6pt

\noindent
{\bf Definition 2.2}.  Given a unimodular code $x$ of length $N$, the autocorrelation function 
(${\acf}$) of $x$ is defined as the sequence of length $2N-1$ 
\[{\acf}_x = x\ast \overline{x}\]
where $\ast$ represents aperiodic convolution and $\overline{x}$ means reversal of $x$.  The elements 
${\acf}_x(k)$ for $k=1-N,\ldots,-1,0,1,\ldots, N-1$ may be written explicitly 
as sums of pairwise products of the elements of $x$:  
\begin{equation}
{\acf}_x(k)=\sum_{i=1}^{N-k} {x[i]}\overline{x[i+k]},
\end{equation}
for $k = 0,1, \ldots, N-1$, where $x[i]$ denotes the $i$-th component of $x$ and $\overline{x[i]}$ represents complex conjugation.  if $k=1-N,\ldots, -1$, then
\[ {\acf}_x(-k) = \overline{{\acf}}_x(k) .\]
\begin{itemize}
\item{} $|{\acf}_x(N-1)| = |x[1]\overline{x[N]}| = 1$.
\item{} When $k=0$, ${\acf}_x(k)$ represents the peak of the autocorrelation, which equals 
\[x_[1]\overline{x[1]} + \ldots + x[N]\overline{x[N]} = ||x||^2=N.\]
\end{itemize}

\vskip 6pt

\noindent
{\bf Definition 2.3}. [5]
A $p$-phase $N \times K$ matrix $Q$ consisting of columns $( x_0,x_1,\ldots ,x_{K-1})$ is said to be a {\it complementary code matrix} if 
\[
{\acf}_{x_0}(n)+{\acf}_{x_1}(n)+\ldots+{\acf}_{x_{K-1}}(n)=NK\delta_n
\]
for $n=-(N-1),\ldots ,-1,0,1,\ldots,(N-1)$ where $\delta_n$ is the Kronecker delta function.

\vskip 6pt

\noindent
{\bf Lemma 2.4}. 
Let $Q=( x_0,x_1,\ldots ,x_{K-1})$ be a $p$-phase $N\times K$ CCM.  Then
\begin{eqnarray}
\sum_{i=0}^{K-1}X_i(z)\tilde{X}_i(z) &=&|X_0(z)|^2+\ldots +|X_{K-1}(z)|^2\nonumber \\
&=&NK  \nonumber
\end{eqnarray}

\section{Doppler Shift in Radar}

Let $T = (x_0,x_1,\ldots,x_{L-1})$ be a modulated pulse train whose ambiguity function
is given by 
\begin{equation}
g(k,\theta) = \sum_{n=0}^{L-1}e^{jn\theta} {\acf}_{x_n}(k),
\end{equation}
where $k$ represents range or time delay and $\theta$ represents Doppler-shift-induced
phase advance.   We define the $z$-transform of a code $x$ of length $N$ by
\[
X(z)=x[0]+x[1]z^{-1}+...+x[N-1]z^{-N+1}
\]
Following Pezeshki-Calderbank-Moran-Howard [4], 
the $z$-transform of 
$g(k,\theta)$ becomes
\begin{equation}
G(z,\theta) = \sum_{n=0}^{L-1} e^{jn\theta} |X_n(z)|^2.
\end{equation}
where
\begin{eqnarray}
|X_n(z)|^2 &=& {\acf}_{x_n}(0) + \sum_{k=1}^{N-1}{\acf}_{x_n}(k)z^k + \nonumber\\
& &    \sum_{k=1}^{N-1}\overline{{\acf}_{x_n}}(k)z^{-k}.   \nonumber
\end{eqnarray}

Next, consider the Taylor expansions of $g(k,\theta)$ and $G(z,\theta)$ about $\theta = 0$:
\begin{eqnarray}
g(k,\theta) &=& \sum_{m=0}^{\infty} c_m(k){{(j\theta)^m}\over{m!}}   \\
G(z,\theta) &= & \sum_{m=0}^{\infty} C_m(z){{(j\theta)^m}\over{m!}}.
\end{eqnarray}
Here, the Taylor coefficients $c_n(k)$ and $C_n(k)$ are given by
\begin{eqnarray}
c_m(k) &=& \sum_{n=0}^{L-1} n^m {\acf}_{x_n}(k)   \\
C_m(z) &= & \sum_{n=0}^{L-1} n^m |X_n(z)|^2.
\end{eqnarray}

The following theorem demonstrates an equivalence in terms of the ``vanishing'' of the Taylor coefficients
$c_m(k)$ and $C_m(z)$.

\vskip 6pt

\noindent
{\bf Theorem 3.1}.  Let $m$ be a non-negative integer. Then $c_m(k) = 0$ for all non-zero $k$ if and only
if $C_m(z)$ is constant and independent of $z$.

\vskip 6pt

\noindent
{\it Proof}.  Assume $c_m(k) = 0$ for all non-zero $k$.  It follows from (2) that
\begin{eqnarray*}
C_m(z) &=&  \sum_{n=0}^{L-1}  n^m|X_n(z)|^2 \nonumber \\
&=&  \sum_{n=0}^{L-1} n^m({\acf}_{x_n}(0) + \sum_{k=1}^{N-1} {\acf}_{x_n}(k)z^k  + \sum_{k=1}^{N-1} \overline{{\acf}_{x_n}}(k)z^{-k})   \nonumber \\
&=&   \sum_{n=0}^{L-1} n^m {\acf}_{x_n}(0) +\sum_{n=0}^{L-1}n^m \sum_{k=1}^{N-1} {\acf}_{x_n}(k)z^k  + \sum_{n=0}^{L-1} n^m \sum_{k=1}^{N-1} \overline{{\acf}_{x_n}}(k)z^{-k} \nonumber \\
&=&  \sum_{n=0}^{L-1} n^m {\acf}_{x_n}(0) + \sum_{k=1}^{N-1}\sum_{n=0}^{L-1}n^m{\acf}_{x_n}(k)z^k + \sum_{k=1}^{N-1}\sum_{n=0}^{L-1} n^m \overline{{\acf}_{x_n}}(k)z^{-k} \nonumber \\
&=&  \sum_{n=0}^{L-1}  n^m  {\acf}_{x_n}(0).  \nonumber
\end{eqnarray*}
This proves that $C_m(z)$ is constant and independent of $z$.  Conversely, assume $C_m(z)$ is constant and
independent of $z$.  Then from the previous calculation we have
\[ C_m(z) = \sum_{n=0}^{L-1} n^m {\acf}_{x_n}(0) + \sum_{k=1}^{N-1}c_m(k)z^k + \sum_{k=1}^{N-1}\overline{c_m}(k)z^{-k}.\]
It follows that $c_m(k) = 0$ for all non-zero $k$ since $C_m(z)$ is independent of $z$.  $\qed$

\section{Generalized Prouhet-Thue-Morse Sequences}

Denote by $S(L) = \{0,1,\ldots,L-1\}$  to be the set consisting of the first $L$
non-negative integers.

\vskip 6pt

\noindent
{\bf Definition 4.1}.  Let $n=n_1n_2\dots n_k$ be the base-$p$ representation of a non-negative integer $n$,
where $n_i \in \{0,1,\ldots,p-1\}$ for $i=1,\ldots,k$.  We
define $v_p(n)\in \mathbb{Z}_p$ to be the least positive residue of the sum of the digits $n_i$ modulo $p$, that is,
\[  v_p(n) \equiv \left(\sum_{i=1}^k n_i\right)   \modul p.\]

\vskip 6pt

\noindent
Note that $v_p(n) = n$ if $0\le n < p$.

\vskip 6pt

\noindent
{\bf Definition 4.2} ([11]).  Let $p$ be a positive integer.  We define the mod-$p$
{\it Prouhet-Thue-Morse}  (PTM) {\it sequence}  $P = \{a_0,a_1,\ldots \}$ to be such that
\[ a_n = v_p(n).\]

\vskip 6pt

\noindent
{\bf Example 4.3}:  Examples of $P$ for $p=2,3,4$ are given below.  Observe that for $p=2$, $P$ reduces to the classical Prouhet-Thue-Morse sequence [13].

\vskip 6pt

\noindent
$p=2$:

\noindent
$P= \{0,1,1,0,1,0,0,1,1,0,0,1,0,1,1,0,\ldots \}$

\vskip 6pt

\noindent
$p=3$:

\noindent
$P = \{0,1,2,1,2,0,2,0,1,1,2,0,2,0,1,0,1,2,\ldots\}$
 
 \vskip 6pt

\noindent
$p=4$:

\noindent
$P = \{0,1,2,3,1,2,3,0,2,3,0,1,3,0,1,2,1,2,3,0,\ldots \}$

\vskip 10pt

\noindent
{\bf Definition 4.4}.  Let $p$ and $M$ be positive integers and set $L = p^{M+1}$.  We define
$\{S_0,S_1,\ldots,S_{p-1}\}$ to be a Prouhet-Thue-Morse (PTM) $p$-block partition
of $S(L) = \{0,1,\ldots,L-1\}$ as follows:  if $v_p(n) = i$, then 
\[ n \in S_i.\]

\vskip 6pt

\noindent
{\bf Example 4.5}: Examples of PTM block partitions are given below.

\noindent
$p=2$, $M=3$, $L=16$:
\begin{eqnarray}
S_0 &=&  \{0,3,5,6,9,10,12,15\} \nonumber  \\
S_1 &=&  \{1,2,4,7,8,11,13,14\}  \nonumber
\end{eqnarray}

\vskip 6pt

\noindent
$p=3$,$M=2$, $L=27$:
\begin{eqnarray}
S_0 &=& \{0,5,7,11,13,15,19,21,26\} \nonumber  \\
S_1 &=&  \{ 1,3,8,9,14,16,20,22,24\}  \nonumber \\
S_2 &=& \{2,4,6,10,12,17,18,23,25\}  \nonumber
\end{eqnarray}

\vskip 6pt

\noindent
$p=4$,$M=2$, $L=64$:
\begin{eqnarray}
S_0 &=& \{0,7,10,13,19,22,25,28,34,37,40,47,49,52,59,62\} \nonumber  \\
S_1 &=&  \{1,4,11,14,16,23,26,29,35,38,41,44,50,53,56,63\}  \nonumber \\
S_2 &=& \{2,5,8,15,17,20,27,30,32,39,42,45,51,54,57,60\}  \nonumber \\
S_3 &=& \{3,6,9,12,18,21,24,31,33,36,43,46,48,55,58,61\}  \nonumber
\end{eqnarray}

\vskip 6pt

\noindent
{\bf Theorem 4.6}  ([11][15],[16]).  Let $p$ and $M$ be positive integers and
set $L=p^{M+1}$.  Define $\{S_0,S_1,\ldots,S_{p-1}\}$ to be a PTM
$p$-block partition of $S(L)=\{0,1,\ldots,L-1\}$.  Then
\[ \sum_{n\in S_0} n^m = \sum_{n\in S_1} n^m = \ldots = \sum_{n\in S_{p-1}} n^m \]
for $m = 1,\ldots, M$.

\vskip 6pt

\noindent
It will be convenient to define $P_m := P_m(p,M) = \sum_{n\in S_0}n^m$ to be
the $m^{th}$ Prouhet sum corresponding to $p$ and $M$.

Let $(A_0, A_1, \ldots )$ be a sequence of elements satisfying the aperiodic property
\[  A_n = A_{v_p(n)}.  \]
We shall define an orthogonal set of sequences $w_i(n)$ whose values are given by the
Rademacher functions [14].  These sequences will be used to define a transformation of
the elements $(A_0, A_1, \ldots, A_{p-1})$ whose invertibility provides a useful
decomposition for isolating sidelobes in the total autocorrelation of a train of coded pulses.

\vskip 6pt

\noindent
{\bf Definition 4.7}.  Let 
\[i = d_{p-1}^{(i)}2^{p-1} + d_{p-2}^{(i)}2^{p-2}+ \ldots + d_1^{(i)}2^1 + d_0^{(i)}2^0\]
be the binary expansion of $i$, where
$i$ is a non-negative integer with $0 \leq i \leq 2^p-1$.
Define $w_0(n), w_1(n), \ldots, w_{2^p-1}(n)$ to be binary $\pm 1$-sequences
\[
w_i(n) = (-1)^{d_{p-1-v_p(n)}^{(i)}}
\]
for $n=0,1,\ldots$.

\vskip 6pt

\noindent
{\bf Theorem 4.8} ([14]).  Define
\[  B_i = \sum_{n=0}^{p-1}w_i(n)A_n \]
for $i = 0,1,\ldots,2^{p}-1$.  Then
\[ A_n = \left({1/2^{p-1}}\right)\sum_{i=0}^{2^{p-1}-1}w_i(n)B_i\]
for $n=0,1,\ldots $.

\vskip 5pt

Because of Theorem 4.8, we shall call $w_0(n),w_1(n),\ldots,w_{2^{p-1}-1}(n)$ the
PTM weights of $A_n$ with respect to $(B_0,B_1,\ldots,B_{2^{p-1}-1})$.

\vskip 5pt

\noindent
{\bf Example 4.9}: Examples illustrating Theorem 4.8 are given below.

\vskip 5pt

\noindent
(1)  $p=2$:  \begin{eqnarray}
B_0=A_0 + A_1, & A_0=\frac{1}{2}(B_0+B_1) \nonumber  \\
B_1 = A_0 - A_1, & A_1=\frac{1}{2}(B_0-B_1) \nonumber
\end{eqnarray}

\vskip 5pt

\noindent
(2)  $p=3$:  \begin{eqnarray}
B_0 = A_0 + A_1  + A_2, & A_0= \frac{1}{4}(B_0+B_1+B_2+B_3)  \nonumber  \\
B_1 = A_0 + A_1 - A_2, & A_1= \frac{1}{4}(B_0+B_1-B_2-B_3)  \nonumber  \\
B_2 =  A_0 - A_1 + A_2, & A_2= \frac{1}{4}(B_0-B_1+B_2-B_3)   \nonumber \\
B_3 =  A_0 - A_1  - A_2  &  \nonumber
\end{eqnarray}

\vskip 6pt

\noindent
{\bf Theorem 4.10} ([14]).  Suppose $L = p^{M+1}$ where $M$ is a non-negative integer.
Write
\begin{equation}
 A_n = (1/2^{p-1})w_0(n)B_0 + (1/2^{p-1})S_p(n)
 \end{equation}
where
\[ S_p(n) = \sum_{i=1}^{2^{p-1}-1}w_i(n)B_i.\]
Then
\begin{equation}
 \sum_{n=0}^{L-1} n^m S_p(n) = N_mB_0
 \end{equation}
for $m=1,\ldots,M$ where
\[  N_m = 2^{p-1}P_m - \sum_{n=0}^{L-1}n^m.\]

\vskip 6pt

\section{Doppler-Tolerant CCM Waveforms}

In this section we generalize the results in [7] and [10] by constructing Doppler-tolerant
CCM waveforms.

\vskip 6pt

\noindent
{\bf Definition 5.1}.  We define a mod-$p$ Prouhet-Thue-Morse (PTM) pulse train
$T = (x_0,x_1,\ldots,x_{L-1})$ to be a sequence satisfying
\[ x_n = x_{v_p(n)}.\]

\vskip 6pt

Let $A_n(k)$ represent sidelobe $k$ for the autocorrelation ${\acf}_{x_n}$ of
code $x_n$.  It follows that $A_n(k) = A_{v_p(n)}(k)$.
At times, the sidelobe index $k$ will be suppressed, when the property being
discussed applies regardless of the particular sidelobe.

We now use the results from the previous section to isolate the sidelobe term given by (9) in the 
ambiguity function $g(k,\theta)$.  Suppose $L = p^{M+1}$ where $M$ is a non-negative
integer.  It follows from (2) and (8) that
\begin{eqnarray}
g_p(\theta) &:=& g(k,\theta) \nonumber \\
&=& \sum_{n=0}^{L-1} A_{v_p(n)}e^{jn\theta} \nonumber \\
&=& \sum_{n=0}^{L-1} ((1/2^{p-1})w_0(n)B_0 + (1/2^{p-1})S_p(n)) e^{jn\theta} \nonumber \\
&=& (1/2^{p-1})B_0\sum_{n=0}^{L-1}e^{jn\theta} + (1/2^{p-1})\sum_{n=0}^{L-1}S_p(n)e^{jn\theta}. \nonumber
\end{eqnarray}
The argument uses the fact that $w_0(n)=1$.  
\vskip 6pt

\noindent
{\bf Example 5.2}.  Let $p = 2$.  Then $g_2(\theta)$ reduces to equation (11) in [7]:
\begin{eqnarray}
g_2(\theta) &=& (1/2)B_0\sum_{n=0}^{L-1}e^{jn\theta} + (1/2)\sum_{n=0}^{L-1}S_2(n)e^{jn\theta} \nonumber \\
&=& (1/2)(A_0+A_1)\sum_{n=0}^{L-1}e^{jn\theta} +  (1/2)(A_0-A_1)\sum_{n=0}^{L-1}w_1(n)e^{jn\theta},\nonumber
\end{eqnarray}
where $w_1(n)=p_n$ is the classical Prouhet-Thue-Morse sequence defined by the
recurrence $p_0=1$, $p(2n)=p(n)$, and $p(2n+1) = -p(n)$.

Define
\[ h_p(\theta) = (1/2^{p-1})\sum_{n=0}^{L-1} S_p(n)e^{jn\theta}\]
so that
\[ g_p(\theta) = (1/2)B_0\sum_{n=0}^{L-1}e^{jn\theta} + h_p(\theta).\]

If $Q = (x_0,x_1,\ldots,x_{K-1})$ is a unimodular $N\times K$ CCM, then $h_p(\theta)$ represents the
sidelobes of $g_p(\theta)$ since $B_0 = A_0+A_1+\ldots + A_{K-1}$ vanishes for all non-zero $k$, being the
sum of the autocorrelation functions of $x_0,x_1,\ldots,x_{K-1}$.  Expanding $h_p(\theta)$ in a Taylor series
about $\theta = 0$:
\[ h_p(\theta) = (1/2^{p-1})\sum_{m=0}^{\infty} s_m\left((j\theta)^m/m!\right)\]
where
\[ s_m = \sum_{n=0}^{L-1}n^m S_p(n).\]

\vskip 6pt

The following result generalizes Theorem 2 in [7].

\vskip 6pt

\noindent
{\bf Theorem 5.3}.  Let $Q$ be a unimodular $N\times K$ CCM consisting of columns 
$(x_0,x_1,\ldots,x_{K-1})$ and
$M$ a positive integer.  Set $L = K^{M+1}$ and extend $Q$ to a pulse train $T=(x_0,x_1,\ldots,x_{K-1},x_K,\ldots,x_{L-1})$ where
\[x_n = x_{v_K(n)}\]
for all $n=0,1,\ldots,L-1$.  Then the Taylor coefficients $s_m$ of $h_K(\theta)$ vanish up to order $M$, namely
\[ s_m = 0\]
for $m = 1,\ldots,M$.

\vskip 6pt

\noindent
{\it Proof}.  Set $p=k$.  It follows from (9) that
\begin{eqnarray}
s_m &=& N_mB_0 \nonumber \\
&=& N_m(A_0+A_1+\ldots + A_{K-1}) \nonumber \\
&=& N_m({\acf}_{x_0}(k) + {\acf}_{x_1}(k) + \ldots + {\acf}_{x_{K-1}}(k) \nonumber \\
&=& 0 \nonumber
\end{eqnarray}
for all non-zero $k$.   $\qed$

\vskip 6pt

Next, we move to the $z$-domain and prove an equivalent version of Theorem 5.3 by generalizing
Theorem 2 in [8], which constructs Doppler-tolerant pulse trains in the $z$-domain.

\vskip 6pt

\noindent
{\bf Theorem 5.4}.  Let $Q$ be a unimodular $N\times K$ CCM consisting of columns
$(x_0,x_1,\ldots,x_{K-1})$ and $M$ a positive integer.  Set $L=K^{M+1}$ and extend $Q$ to
a pulse train $T=(x_0,x_1,\ldots,x_{K-1},x_K,\ldots,x_{L-1})$ where
\[ x_n = x_{v_K(n)}\]
for all $n=0,1,\ldots,L-1$.  Then the Taylor coefficients $C_m(z)$ are independent of $z$ up to
order $M$, namely
\[ C_m(z) = NKP_m\]
for $m=1,\ldots,M$ where $P_m$ is the $m^{th}$ Prouhet sum corresponding to $K$ and $M$.

As in [4], we call $T$ a mod-$K$ Prouhet-Thue-Morse (PTM) pulse train
of length $L$.  

\vskip 6pt

\noindent
{\it Proof}.  Let $\{S_0,S_1,\ldots,S_{K-1}\}$ be a PTM $K$-block partition of $S=\{0,1,\ldots,L-1\}$.
It follows from Theorem 4.6 and Lemma 2.4 that
\begin{eqnarray}
C_m(z) &=&  \sum_{n=0}^{L-1}n^m|X_n(z)|^2  \nonumber   \\
&=& \sum_{n\in S_0}n^m|X_{v_K(n)}(z)|^2 + \sum_{n\in S_1}n^m|X_{v_K(n)}(z)|^2  \nonumber \\
& &  + \ldots + \sum_{n\in S_{K-1}}n^m|X_{v_K(n)}(z)|^2  \nonumber \\
&=& |X_0(z)|^2\sum_{n\in S_0}n^m + |X_1(z)|^2\sum_{n\in S_1}n^m  \nonumber \\
& &  + \ldots + |X_{K-1}(z)|^2\sum_{n\in S_{K-1}}n^m   \nonumber \\
&=& (|X_0(z)|^2 + |X_1(z)|^2  + \ldots + |X_{K-1}(z)|^2)P_m   \nonumber \\
&=& NKP_m   \nonumber 
\end{eqnarray}
for $m=1,2,\ldots,M$.  $\qed$

\vskip 6pt

\noindent 
{\bf Example 5.5}: Examples of PTM pulse trains are given below.

\noindent 
1. Let $K=2$, $M=3$, and $(x_0,x_1)$ be a binary $N\times 2$ CCM (Golay pair).  Then the following is a mod-2 PTM pulse train of length $L=2^4=16$: 
\[T=(x_0,x_1,x_1,x_0,x_1,x_0,x_0,x_1,x_1,x_0,x_0,x_1,x_0,x_1,x_1,x_0)
\]

\vskip 5pt

\noindent 
2. Let $K=3$, $M=2$, and $(x_0,x_1,x_2)$ be a tri-phase $N\times 3$ CCM.  Then the following is a mod-3 PTM pulse train of length $L=3^3=27$: 
\begin{eqnarray}
T&=&(x_0, x_1, x_2, x_1, x_2, x_0, x_2, x_0, x_1, x_1, x_2, x_0, x_2, x_0, \nonumber \\
& &x_1, x_0, x_1, x_2, x_2, x_0, x_1, x_0, x_1, x_2, x_1, x_2, x_0)\nonumber
\end{eqnarray}

\vskip 5pt
\noindent 
3. Let $K=4$, $M=2$, and $(x_0,x_1,x_2,x_3)$ be a unimodular $N\times 4$ CCM.  Then the following is a mod-4 PTM pulse train of length $L=4^3=64$: 
\begin{eqnarray}
T&=& (x_0, x_1, x_2, x_3,  x_1, x_2, x_3, x_0,  x_2, x_3, x_0, x_1,  \nonumber \\
& & x_3, x_0, x_1, x_2, x_1, x_2, x_3, x_0,x_2, x_3, x_0, x_1, \nonumber \\
& & x_3, x_0, x_1, x_2, x_0, x_1, x_2, x_3, x_2, x_3, x_0, x_1, \nonumber \\
& & x_3, x_0, x_1, x_2, x_0, x_1, x_2, x_3, x_1, x_2, x_3, x_0, \nonumber \\
& & x_3, x_0, x_1, x_2,x_0, x_1, x_2, x_3, x_1, x_2, x_3, x_0, \nonumber \\
& & x_2, x_3, x_0, x_1) \nonumber 
\end{eqnarray}

\section{ESP Staggered Pulse Trains}

In this section we introduce pulse trains, called ESP staggered pulse trains, that
provide the same Doppler tolerance as PTM pulse trains but are
generally shorter in length, by using multiple antennas to transmit separate
pulse trains staggered in time.  We begin with definitions of delayed pulse trains and partitions
of arbitrary sets of non-negative integers (not necessarily consecutive as with PTM partitions) having equal sums of powers.

\vskip 5pt

\noindent
{\bf Definition 6.1}. We define a {\em delayed} pulse train
\[
T(d)=(x_0,x_1,...,x_{L-1})
\]
of length $L$ as one having a delay of $d$ pulses in the sense that its ambiguity function has the form
\[
g_T(k,\theta,d)=\sum_{n=0}^{L-1} {\acf}_{x_n}e^{i(n+d)\theta}
\]

\noindent
{\bf Definition 6.2}.  Let $S$ be a set of non-negative integers and $P = \{S_0,S_1,\ldots,S_{p-1}\}$ be a $p$-block partition
of $S$.  We say that $P$ has {\it equal sums of (like) powers (ESP)} of degree $M$ if
\[  \sum_{n\in S_0} n^m = \sum_{n\in S_1} n^m = \ldots = \sum_{n\in S_{p-1}} n^m\]
for $m=1,\ldots,M$.  In that case, we define 
\[P_m := P_m(C) = \sum_{n\in S_0} n^m.\]

The following examples demonstrates our concept of using MIMO (multiple-input multiple-output) radar to transmit ESP pulse trains whose overall transmission period is shorter than PTM pulse trains.

\vskip 5pt

\noindent {\bf Example 6.3}: (Second-order nulls) Let $S=\{0,1,2,4,5,6\}$ and consider the 2-block partition $\mathcal{P}=(S_0,S_1)$ of $S$, where 
$S_0 = (0, 4, 5)$ and $S_1 = (1, 2, 6)$.  Then $\mathcal{P}$ has ESP of degree 2 since
\begin{align*}
0+4+5 & =1+2+6 \\
0^2+4^2+5^2 & =1^2+2^2+6^2
\end{align*}
Observe that this partition consists of only six values (skipping the value 3) and is smaller in size than the 2-block PTM partition of $\{0,1,...,7\}$.  Then given a Golay pair of codes $(x_0,x_1)$, we can of course construct a single pulse train based on the partition above by inserting a gap or fill pulse for the value at position 3:
\[
T=(x_0,x_1,x_1, \_\_ ,x_0,x_0,x_1)
\]
This approach however is impractical in terms of transmission.  On the other hand, we can modify the partition $\mathcal{P}$ so that it includes the value 3 in both sets:
\begin{align*}
S_0 & =(0,3,4,5) \\
S_1 & =(1,2,3,6)
\end{align*}
Note that $\mathcal{P}$ is no longer a collection of mutually disjoint sets but continues to have ESP of degree 2.  Suppose we then transmit two separate pulse trains of length 4, $T_0$ and $T_1$ (each from a separate antenna), but staggered in the sense that we delay the transmission of $T_1$ by 3 pulses as follows:
\begin{align*}
T_0 & =(x_0, x_1, x_1, x_0) \\
T_1(3) & = \hspace{44pt} (x_1, x_0,x_0,x_1)
\end{align*}
Here, $T_0$ transmits pulses corresponding to the first two values of $S_0$ (positions 0 and 3) and the first two values of $S_1$ (positions 1 and 2).  Similarly for $T_1(3)$, but corresponding to the last two values of $S_0$ and $S_1$.  If we sum the composite ACFs of both pulse trains, then we obtain
\begin{align*}
g(k,\theta) & =g_{T_0}(k,\theta)+g_{T_1}(k,\theta,3) \\
& = {\acf}_{x_0}(k)+{\acf}_{x_1}(k)e^{i\theta}+{\acf}_{x_1}(k)e^{2i\theta} \\
& \ \ \ \ + ({\acf}_{x_0}(k)+{\acf}_{x_1}(k))e^{3i\theta}+{\acf}_{x_0}(k)e^{4i\theta} \\
& \ \ \ \ +{\acf}_{x_0}(k)e^{5i\theta}+{\acf}_{x_1}(k)e^{6i\theta}
\end{align*}
To show that $g(k,\theta)$ has Doppler nulls of order 2 at $\theta=0$, we compute its Doppler (Taylor) coefficients:
\begin{align*}
c_m(k) & = g^{(m)}(k,0) \\
& = (0^m  + 3^m +4^m +5^m){\acf}_{x_0}(k) \\
& \ \ \ \ +(1^m +2^m +3^m +6^m) {\acf}_{x_1}(k) \\
& = P_m ({\acf}_{x_0}(k)+{\acf}_{x_1}(k)) \\
& = 2NP_m\delta_{k} 
\end{align*}
for $m=0,1,2$.   This demonstrates that we can achieve the same Doppler tolerance as with a single PTM pulse train of length 8 by using instead two staggered (but overlapping) pulse trains of length 4 to reduce the total transmission time from 8 pulses down to 7 pulses.  Note however that the total number of pulses transmitted is the same, namely 8, in both cases.

\vskip 5pt

\noindent {\bf Example 6.4}: (Third-order nulls) Consider the following 2-block partition $\mathcal{P}=(S_0,S_1)$, where
\begin{align*}
S_0 & = (0, 4, 7, 11) \\
S_1 & = (1, 2, 9, 10)
\end{align*}
 which has ESP of degree 3, namely
 \[
 0^m+4^m+7^m+11^m=1^m+2^m+9^m+10^m
 \]
 for $m=0,1,2,3$.  As in the previous example, we modify this partition so that both sets $S_0$ and $S_1$ contain each of the values 3, 5, 6, and 8:
\begin{align*}
S_0 & =(0,3,4,5,6,7,8,11) \\
S_1 & =(1,2,3,5,6,8,9,10)
\end{align*}
We now transmit four pulse trains $T_0$, $T_1(3)$, $T_2(5)$, $T_3(8)$ on separate antennas having delays $0, 3, 5, 8$, respectively:
\begin{align*}
T_0 & =(x_0, x_1, x_1, x_0) \\
T_1(3) & = (x_1, x_0,x_0,x_1) \\
T_2(5) & =  (x_1, x_0,x_0,x_1) \\
T_3(8) & =  (x_0, x_1,x_1,x_0) 
\end{align*}
Then it can be shown that the Doppler coefficients of the composite ambiguity function $g(k,\theta)$
has Doppler nulls of order 3:
\begin{align*}
c_m(k) & = (0^m  + 3^m +4^m +5^m+6^m+7^m+8^m \\
& \ \ \ \ +11^m){\acf}_{x_0}(k)+(1^m +2^m +3^m +5^m \\
& \ \ \ \ + 6^m+8^m+9^m+10^m) {\acf}_{x_1}(k) \\
& = P_m ({\acf}_{x_0}(k)+{\acf}_{x_1}(k)) \\
& = 2NP_m\delta_{k} 
\end{align*}
for $m=0,1,2,3$.  Thus, we have reduced the total transmission time from 16 pulses (for a single PTM pulse train of length 16 having the same Doppler tolerance) down to 12 by using instead four pulse trains transmitted separately.  Again, note that the total number of pulses transmitted is the same (16) in both cases.

\vskip 5pt

\noindent {\bf Example 6.5}: (Fifth-order nulls) Consider the following 2-block partition $\mathcal{P}=(S_0,S_1)$ which has ESP of degree 5:
\begin{align*}
S_0 & = (0, 5,6,16,17,22) \\
S_1 & = (1, 2,10,12,20,21)
\end{align*}
We again modify this partition to include the values $\{3, 4, 7, 8, 9, 11, 13, 14, 15, 18, 19\}$ without changing its degree:
\begin{align*}
S_0 & =(0,3,4,5,6,7,8,9,11,13,14,15,16,17,18,19,22) \\
S_1 & =(1,2,3,4,7,8,9,10,11,12,13,14,15,18,19,20,21)
\end{align*}
We then transmit seven pulse trains $T_0$, $T_1(3)$, $T_2(7)$, $T_3(8)$, $T_4(10)$, and $T_5(13)$, and $T_6(18)$ having delays 0, 3, 7, 8, 10, 13, and 18, respectively:
\begin{align*}
T_0 & = (x_0, x_1, x_1, x_1, x_1) \\
T_1(3) & = (x_0,x_0,x_0,x_0,x_1) \\
T_2(7) & = (x_0,x_0,x_0) \\
T_3(8) & = (x_1,x_1,x_1,x_1,x_1) \\
T_4(10) &= (x_0,x_1,x_1,x_1,x_1) \\
T_5(13) & = (x_0,x_0,x_0,x_0,x_0,x_0,x_0) \\
T_6(18) &= (x_1,x_1,x_1,x_1,x_0)
\end{align*}
Again it can be shown that the Doppler coefficients of the composite ambiguity function $g(k,\theta)$ has Doppler nulls of order 5.  Thus, we have reduced the total transmission time from 64 pulses (for a single PTM pulse train of length 64 having the same Doppler tolerance) down to 23 by using instead seven pulse trains transmitted by separate antennas.  Unlike Examples 6.3 and 6.4, the total number of pulses transmitted for all seven staggered pulse trains is only 40 in comparison to 64 for a single PTM pulse train.  We observe that the three pulse trains $T_2(7)$, $T_3(8)$, and $T_5(13)$ are constant in value.

 \section{Conclusions}
 
 Pezeshki, Calderbank, Howard, and Moran have shown that Doppler tolerance can be achieved in match-filtered 
 trains of time-separated pulses encoded with Golay complementary pairs.  The key is to map the two codes to the pulses 
 in the train using the well-known Thue-Morse sequence.  Depending on the number of pulses that can be supported 
 for a particular application, the Doppler tolerance can be achieved to any desired order.   This paper has shown that 
 the same is possible with complementary code sets containing more than two codes.  Generalization is achieved by 
 exploiting several number-theoretic concepts, including equal sums of like powers, the digit sum function, 
 and the generalization to $m\ge 2$ symbols of the classical  two-symbol Thue-Morse sequence.  In addition, it is shown that certain ESP pulse trains having shorter lengths than PTM pulse trains can be used to obtain the same Doppler tolerance by employing multiple antennas to transmit these pulse trains staggered in time.
 
 \section{References}
 
 \noindent
[1]  Levanon, N. and Mozeson, E., {\it Radar Signals}, Wiley, NY, 2005.

\vskip 6pt

\noindent
[2] Van Nee, R., ``OFDM codes for peak-to-average power reduction and error correction,'' {\it Proceedings of GLOBECOM 96}, vol. 1, pp. 740-744,
Nov. 1996.

\vskip 6pt

\noindent
[3] Davis, J. and Jedwab, J., ``Peak-to-mean power control and error control OFDM, Golay complementary sequences and
Reed Muller codes,'' {\it IEEE Transactions on Information Theory}, vol. 45, no. 7, pp. 2397 - 2417, November 1997.

\vskip 6pt

\noindent
[4]  Golay, M.J.E., ``Complementary Series,''  {\it IEEE Transactions on Information Theory}, vol. 7, pp. 82-87, April 1961.

\vskip 6pt

\noindent
[5]  Coxson, G.E. and Haloupek, W., ``Construction of complementary code matrices for waveform
design,''  {\it IEEE Transactions on Aerospace and Electronic Systems}, vol. 49 (2013), no. 3, pp. 1806-1816.

\vskip 6pt

\noindent
[6]  Ducoff, M.R. and Tietjen, B.W.,``Pulse compression radar,'' Chapter 8 in Skolnik, M., {\it Radar Handbook}, 3rd Ed.,
McGraw-Hill, 2008. 

\vskip 6pt

\noindent
[7]  Chi, Y., Pezeshki, A. and Calderbank, A.R., ``Complementary waveforms for sidelobe suppression
and  radar polarimetry,''  in {\it Principles of Waveform Diversity and Design}, M. Wicks, E. Mokole,
S. Blunt, R. Schneible and V. Amuso, editors, SciTech, Raleigh, NC, 2011.

\vskip 6pt

\noindent
[8]  Pezeshki, A., Calderbank, A.R., Moran, W. and Howard, S.D., ``Doppler resilient
Golay complementary waveforms,''  {\it IEEE Transactions on Information Theory}, vol. 54, no. 9, pp. 4254-4266, Sept. 2008.

\vskip 6pt

\noindent
[9]  Searle, S.J., Howard, S.D., and Moran, W., ``On the formation of
composite ambiguity functions with frequency separated Golay coded pulses,'' {\it IEEE Transactions on
Aerospace and Electronic Systems}, vol. 45, no. 4, pp. 1580-1597, Jan. 2009.

\vskip 6pt

\noindent
[10] Searle, S.J., Howard, S.D., and Moran, W., ``Nonlinear complementary waveform sets for clutter
suppression,'' in {\it Principles of Waveform Diversity and Design}, M. Wicks, E. Mokole,
S. Blunt, R. Schneible and V. Amuso, editors, SciTech, Raleigh, NC, 2010, pp. 772-800.

\vskip 6pt

\noindent
[11]  Wright, E.M., ``Prouhet's 1851 solution of the Tarry-Escott problem of 1910,'' {\it American Mathematical Monthly}, vol. 102 (1959), pp. 199-210.

\vskip 6pt

 \noindent
 [12]  Allouche, J.-P. and Shallit, J., ``Sum of digits, overlaps, and palindromes,''  {it Discrete Mathematics
 and Theoretical Computer Science}, vol. 4, pp. 1-10, 2000.

\vskip 6pt

\noindent
 [13] Allouche, J.-P. and Shallit, J., ``The ubiquitous Prouhet-Thue-Morse sequence,'' {\it Sequences and Their applications, Proc. SETA'98}, C. Ding, T. Helleseth, and H. Niederreiter, eds.. New York: Springer-Verlag, pp. 1-16, 1999.
 
 \vskip 6pt

\noindent
[14]  Nguyen, H.D.,  ``A mixing of Prouhet-Thue-Morse sequences and Rademacher functions,'' preprint, 2014: arXiv:1405.6958.

\vskip 6pt

\noindent
[15] Prouh\`{e}t, E., ``M\'{e}moires sur quelques relations entre les puissances des nombres,''  {\it C.R. Acad. Sci.}, Paris, vol. 33, p. 225, 1851.

\vskip 6pt

\noindent
[16] Lehmer, D.H., ``The Tarry-Escott problem,''  {\it Scripta Math.}, vol. 13, pp. 37-41, 1947.

\vskip 6pt

\noindent
[17]  Lander, L.J., Parkin, T.R. and Selfridge, J.L., ``A Survey of Equal Sums of Like Powers,''
{\it Mathematics of Computation},  vol. 21, no. 99  pp. 446-459, July 1967.

\end{document}